\documentclass[%
 reprint,
showpacs,preprintnumbers,
 amsmath,amssymb,
 prl,
]{revtex4-1}

\usepackage{graphicx}% Include figure files
\usepackage{dcolumn}% Align table columns on decimal point
\usepackage{bm}% bold math
\def\Vec#1{\mbox{\boldmath $#1$}}
\newcommand{\Prob}{P}
\newcommand{\Trans}{W}
\newcommand{\boxr}{r}

\begin{document}

\preprint{}

\title{Fluctuation Theorem for Hidden Entropy Production}%

\author{Kyogo Kawaguchi}
 \email{kyogok@daisy.phys.s.u-tokyo.ac.jp}
\author{Yohei Nakayama}%
 \email{nakayama@daisy.phys.s.u-tokyo.ac.jp}
\affiliation{%
 Department of Physics, The University of Tokyo, Hongo 7-3-1, Tokyo 113-0033, Japan
}%

\date{\today}% It is always \today, today,

\begin{abstract}
In the general process of eliminating dynamic variables in Markovian models, there exists a difference in the irreversible entropy production between the original and reduced dynamics. We call this difference the hidden entropy production, since it is an invisible quantity when only the reduced system's view is provided. We show that this hidden entropy production obeys a new integral fluctuation theorem for the generic case where all variables are time-reversal invariant, therefore supporting the intuition that entropy production should decrease by coarse graining. It is found, however, that in cases where the condition for our theorem does not hold, entropy production may also increase due to the reduction. The extended multibaker map is investigated as an example for this case.
\end{abstract}

                     % Classification Scheme.
\maketitle

%\tableofcontents

\textit{Introduction.}---
Some of the main studies in statistical physics concern reducing variables from the original microscopic equations to gain knowledge on the effective dynamics.
The reduction is typically justified by the time scale separation between variables, 
and the descriptions of effective kinetics become rigorous in the so-called Markovian limit \cite{spohn1980,vankampen1985}.
Recent studies have focused on how physical quantities may vary according to the descriptions at different scales \cite{sekimoto_book2010}, or are related to each other when defined in many partial systems \cite{sagawa2012a}.

In this paper, we consider how the irreversible (total) entropy production $\Sigma$ varies by the general procedure of eliminating variables.
Here, $\Sigma$ is the sum of the Shannon entropy difference $s$ and the 
heat entropy production $\sigma$.
This $\Sigma$, defined uniquely in a Markovian dynamics setup \cite{lebowitz1999,crooks1999}, has a positive ensemble 
average (denoted by $\langle \cdot \rangle$) value,
\begin{eqnarray}
\langle \Sigma \rangle=\langle s \rangle+\langle \sigma \rangle \geq 0. \label{secondlaw_intro}
\end{eqnarray}
Since the inequality (\ref{secondlaw_intro}) may be considered as the mesoscopic version of the second law of thermodynamics, the scale dependent descriptions of $\Sigma$ is obviously an important matter.

The main purpose of this paper is to clarify the behavior of the difference between $\Sigma$ defined in the original dynamics, and the entropy production $\widetilde{\Sigma}$ defined in the reduced system,
\begin{eqnarray}
\Xi = \Sigma-\widetilde{\Sigma}.
\end{eqnarray}
We call $\Xi$ the hidden entropy production.
It is known in the context of infomation theory \cite{CoverThomas1991} that if $\Sigma \to \widetilde{\Sigma}$ can be written as a reduction of variables in the Kullback-Leibler divergence by the Markov map, $\langle \Xi \rangle$ is larger than zero.
Corresponding to this notion, we first show for the case where all variables in the original dynamics are time-reversal invariant, that $\Xi$ obeys the integral fluctuation theorem, 
\begin{eqnarray}
\left \langle e^{-\Xi} \right \rangle  = 1,\label{integral}
\end{eqnarray}
from which $\langle \Xi \rangle \geq0$ follows directly.

An important situation where $\langle \Xi \rangle \geq0$ should not hold is when a time-reversal symmetric system (e.g. Hamiltonian dynamics) is reduced to an irreversible stochastic dynamics.
We find that (\ref{integral}) does not hold if and only if the original dynamics includes time-reversal anti-symmetric variables and the symmetry of the density function is broken for them.
As our second main result, we demonstrate these points by investigating the extended multibaker model, which is a Hamiltonian-like dynamics that reduces to a simple random walk in the mesoscopic regime.

\textit{Model and definitions.}---
We consider the Markov chain dynamics on a continuous state space.
The continuous variables $x$ and $y$ may each represent many variables,
nevertheless we use a single variable notation.
The time evolution of the probability density function ${\Prob}_t(x,y)$ follows the equation,
\begin{eqnarray}
{\Prob}_{t+\Delta t}(x,y)=\int dx'dy' {\Prob}_{t}(x',y'){\Trans}_{\lambda(t)}(x,y|x',y'), \label{master_eq}
\end{eqnarray}
where $\Delta t$ is the (infinitesimal) time step.
${\Trans}_{\lambda(t)} (x,y|x',y')$ is the transition probability from $(x',y')$ to $(x,y)$ between time $t$ and $t+\Delta t$,
and we assume that it is controlled by the time-dependent external parameter $\lambda(t)$.
The integral by $x'$ and $y'$ in Eq.~(\ref{master_eq}) is taken over the whole space,
and we have set $\int dx dy  {\Trans}_{\lambda(t)}(x,y|x',y') =1$.

Let $\Vec{x}_N = (x_0, x_1,...,x_N)$, $\Vec{y}_N = (y_0, y_1,...,y_N)$ be the stochastic path taken by $(x,y)$ during the $N$ time steps starting from $t=0$.
We define the stochastic Shannon entropy difference assigned to this path as,
\begin{eqnarray}
s(\Vec{x}_N,\Vec{y}_N) := \log \frac{{\Prob}_0(x_0,y_0)}{{\Prob}_{N\Delta t}(x_N,y_N)}. \label{micro_ent}
\end{eqnarray}
Defining the path transition probability as ${\Trans}_{\lambda} (\Vec{x}_N,\Vec{y}_N|x_0,y_0):=\prod_{i=0}^{N-1} {\Trans}_{\lambda(i\Delta t)} (x_{i+1},y_{i+1}|x_i,y_i)$, the path probability as ${\Prob}_\lambda ( \Vec{x}_N,\Vec{y}_N) := {\Prob}_0(x_0,y_0) {\Trans}_{\lambda} (\Vec{x}_N,\Vec{y}_N|x_0,y_0)$, and the Shannon entropy of the whole system at time $t$ as $S(t) :=  -\int dx dy {\Prob}_t(x,y)\log {\Prob}_t(x,y)$, we have $\langle s(\Vec{x}_N,\Vec{y}_N) \rangle_{\lambda, N}=S(N\Delta t) - S(0)$.
Here the bracket $\langle \cdot \rangle_{\lambda, N}$ denotes the average $\int d\Vec{x}_N d\Vec{y}_N {\Prob}_\lambda ( \Vec{x}_N,\Vec{y}_N) \cdot$, where $d\Vec{x}_N:=\prod_{i=0}^N dx_i$ and $d\Vec{y}_N:=\prod_{i=0}^N dy_i$.
Using $\widetilde{\Prob}_t (x):= \int dy {\Prob}_t (x,y)$, we further define the coarse-grained stochastic Shannon 
entropy difference as,
\begin{eqnarray}
\widetilde{s}(\Vec{x}_N) := \log \frac{\widetilde{\Prob}_0(x_0)}{\widetilde{\Prob}_{N\Delta t}(x_N)}. \label{micro_ent_cg}
\end{eqnarray}
The Boltzmann constant is set to 1 throughout the paper.

Next we define the heat entropy production.
The reverse trajectory of $(\Vec{x}_N,\Vec{y}_N)$ is written as $(\Vec{x}^\dagger_N,\Vec{y}^\dagger_N)$, where
$\Vec{x}^\dagger_N = (\bar{x}_N, \bar{x}_{N-1},...,\bar{x}_{0}), \Vec{y}^\dagger_N = (\bar{y}_N, \bar{y}_{N-1},...,\bar{y}_{0})$
with $\bar{x}_i$ being the time reversal of $x_i$.
Now the heat entropy production corresponding to the 
$N$ step trajectory is,
\begin{eqnarray}
\sigma(\Vec{x}_N,\Vec{y}_N) :=  \log \frac{{\Trans}_{\lambda} (\Vec{x}_N,\Vec{y}_N|x_0,y_0)}{{\Trans}_{\lambda^\dagger} (\Vec{x}^\dagger_N,\Vec{y}^\dagger_N| \bar{x}_N,\bar{y}_N)}. \label{micro_heat}
\end{eqnarray}
Here, ${\Trans}_{\lambda^\dagger}$ is the transition probability assigned to the time-reversed protocol, ${\Trans}_{\lambda^\dagger} (\Vec{x}^\dagger_N,\Vec{y}^\dagger_N| \bar{x}_N,\bar{y}_N) := \prod_{i=0}^{N-1} {\Trans}_{\bar{\lambda}((N-i)\Delta t)} (\bar{x}_{i},\bar{y}_{i}|\bar{x}_{i+1},\bar{y}_{i+1})$, which is defined using $\bar{\lambda}(t)$, the time reversal of $\lambda(t)$ [for example if the control is by the magnetic field, $\bar{\lambda}(t)$ corresponds to $\lambda(t)$ with reversed direction].
It is known that $\sigma(\Vec{x}_N,\Vec{y}_N)$ corresponds to the entropy production induced by the energy transfer from the variables $(x,y)$ to the hidden degrees of freedom in general stochastic models \cite{kurchan1998,lebowitz1999}, and in the Hamiltonian system including heat baths \cite{jarzynski2000}.
We note however that in particular Langevin models we introduce in the next section [(\ref{xcg}) and (\ref{xpmodel_cg})], $\sigma(\Vec{x}_N,\Vec{y}_N)$ does not directly correspond to the total heat dissipation \cite{nakayama2012}.
Let the coarse-grained path transition probabilities be,
\begin{eqnarray}
\widetilde{\Trans}_{\lambda} [\Vec{x}_N|x_0,P_0(\cdot)]&& :=  \int d\Vec{y}_N \frac{{\Prob}_\lambda(\Vec{x}_N,\Vec{y}_N)}{\widetilde{\Prob}_0(x_0)} , \label{cond_cg}\\
\widetilde{\Trans}_{\lambda ^\dagger} [\Vec{x}^\dagger_N|\bar{x}_N, P_{N\Delta t}(\cdot)]&& :=  \int d\Vec{y}_N^\dagger \frac{{\Prob}_{\lambda^\dagger}(\Vec{x}_N^\dagger,\Vec{y}_N^\dagger)}{\widetilde{\Prob}_{N\Delta t}(\bar{x}_N)}, \label{cond_cg2}
\end{eqnarray}
where ${\Prob}_{\lambda^\dagger}(\Vec{x}_N^\dagger,\Vec{y}_N^\dagger):={\Prob}_{N\Delta t}(\bar{x}_N,\bar{y}_N) {\Trans}_{\lambda^\dagger} (\Vec{x}^\dagger_N,\Vec{y}^\dagger_N|\bar{x}_N,\bar{y}_N)$.
Then the coarse-grained heat entropy production is defined as,
\begin{eqnarray}
\widetilde{\sigma}(\Vec{x}_N) :=
\log \frac{\widetilde{\Trans}_{\lambda} [\Vec{x}_N|x_0,P_0(\cdot)]}
{\widetilde{\Trans}_{\lambda ^\dagger} [\Vec{x}^\dagger_N|\bar{x}_N, P_{N\Delta t}(\cdot)]}. \label{micro_heat_cg}
\end{eqnarray}
Note that the coarse-grained transition probabilities $\widetilde{\Trans}_{\lambda}, \widetilde{\Trans}_{\lambda 
^\dagger}$ are in general non-Markovian [we omitted the $P_0,P_{N\Delta t}$ dependence in the left hand side of Eq.~(\ref{micro_heat_cg})].

Let the total entropy production be $\Sigma(\Vec{x}_N,\Vec{y}_N):=s(\Vec{x}_N,\Vec{y}_N)+\sigma(\Vec{x}_N,\Vec{y}_N)$, 
the coarse-grained total entropy production be 
$\widetilde{\Sigma}(\Vec{x}_N):=\widetilde{s}(\Vec{x}_N)+\widetilde{\sigma}(\Vec{x}_N)$, and the difference between them be,
\begin{eqnarray}
\Xi(\Vec{x}_N,\Vec{y}_N) := \Sigma(\Vec{x}_N,\Vec{y}_N)-\widetilde{\Sigma}(\Vec{x}_N). \label{def_anom}
\end{eqnarray}
This  $\Xi(\Vec{x}_N,\Vec{y}_N)$ is the entropy production that could not be caught when only the $x$ dynamics is observed, thus we call it the hidden entropy production.
Now for $\Sigma$ and $\widetilde{\Sigma}$ the followings hold,
\begin{eqnarray}
\left \langle e^{-\Sigma\left(\Vec{x}_N,\Vec{y}_N\right)} \right \rangle_{\lambda, N} &=& 1, \label{integral_tot} \\
\left \langle e^{-\widetilde{\Sigma} \left(\Vec{x}_N \right)}\right \rangle_{\lambda, N} &=& 1. \label{integral_cg}
\end{eqnarray}
Equation (\ref{integral_tot}) is the well known integral fluctuation theorem \cite{jarzynski1997,crooks1999}. From this equality and the Jensen's inequality we may show $\langle \Sigma(\Vec{x}_N,\Vec{y}_N) \rangle_{\lambda, N} = \langle s(\Vec{x}_N,\Vec{y}_N) \rangle _{\lambda,N} + \langle \sigma(\Vec{x}_N,\Vec{y}_N) \rangle _{\lambda,N} \ge 0$, the second law corresponding to the whole Markovian dynamics. Eq.~(\ref{integral_cg}) is the
integral fluctuation theorem for the coarse-grained system. The special case of Eq.~(\ref{integral_cg}) was mentioned in \cite{kawai2007}.

\textit{Main theorem.}---
The first main result we present is that for the case where the variables $x,y$ are time-reversal invariant ($x=\bar{x}, y=\bar{y}$), the following equality holds.
\begin{eqnarray}
\left \langle e^{-\Xi \left(\Vec{x}_N,\Vec{y}_N \right)} \right \rangle_{\lambda, N} &=& 1. \label{integral_anom} 
\end{eqnarray}
This will be proved later.
From Eq.~(\ref{integral_anom}) and Jensen's inequality, we see that the hidden entropy production is positive on average,
\begin{eqnarray}
\left \langle \Xi\left(\Vec{x}_N,\Vec{y}_N \right)\right \rangle_{\lambda, N} \geq 0, \label{ineq_anom}
\end{eqnarray}
which means that the entropy production decreases due to the elimination of $y$.

As an example, we introduce a two-dimensional overdamped Langevin model \cite{nakayama2012}.
Taking $\gamma$ as the friction constant, $T$ as the temperature, $\xi(t)$ and $\xi'(t)$ as the independent Gaussian white noises with unit variance, the model is written as,
\begin{eqnarray}
\begin{array}{l}
\gamma \dot{x} = -k_{\lambda(t)} x - f_{\lambda(t)}\alpha y + \sqrt{2\gamma  T_{\lambda(t)}} \xi(t),\\
\gamma \dot{y} = -k_{\lambda(t)}\alpha^2 y + f_{\lambda(t)}x/\alpha + \sqrt{2\gamma T_{\lambda(t)}} \xi '(t).
\end{array} \label{xymodel}
\end{eqnarray}
Here, $k>0$ is the spring constant, $f \neq 0$ is the amplitude of the nonequilibrium force, and $\alpha>0$ is the 
non-dimensional parameter that controls the time scale difference between $x$ and $y$.
In the $\alpha \to \infty$ limit, the fast-moving variable $y$ could be eliminated from Eq.~(\ref{xymodel}) by standard 
singular perturbation methods \cite{vankampen1985}, and we obtain a closed dynamics of $x$, 
\begin{eqnarray}
\gamma \dot{x} &=& -k_{\lambda(t)} x + \sqrt{2\gamma T_{\lambda(t)}} \xi(t). \label{xcg}
\end{eqnarray}
We assumed that the modulation of $k,f,T$ following the parameter $\lambda(t)$ is sufficiently slow compared to 
the time scale of the $y$ dynamics.
Note that in Eq.~(\ref{xcg}), the apparent nonequilibrium force due to $f$ has vanished.
Denoting the stochastic path from time 0 to $\tau$ by $(\Vec{x}_\tau,\Vec{y}_\tau)$, 
Eq.~(\ref{integral_anom}) suggests that
the hidden entropy production $\Xi$ defined by Eq.~(\ref{def_anom}) satisfies $\left \langle \exp[-\Xi(\Vec{x}_\tau,\Vec{y}_\tau)] \right \rangle_{\lambda, \tau} =1$ for any $\alpha$.
In the particular case where $\alpha$ is sufficiently large, the coarse-grained transition probabilities (\ref{cond_cg}) and (\ref{cond_cg2}) become equal to the Markovian transition probability corresponding to Eq.~(\ref{xcg}), and the hidden entropy production may be written as,
\begin{eqnarray}
&&
\Xi(\Vec{x}_\tau, \Vec{y}_\tau) = \log \frac{{\Prob}_0(x_0,y_0)\widetilde{\Prob}_\tau(x_\tau)}{{\Prob}_\tau(x_\tau,y_\tau)\widetilde{\Prob}_0(x_0)}  \label{Xi_xy}\\ 
&& +\int_0 ^\tau \frac{dt}{T_{\lambda(t)}} \left[ -\dot{x} \circ f_{\lambda(t)}\alpha y  -\dot{y} \circ \left(k_{\lambda(t)} \alpha^2 y  - \frac{f_{\lambda(t)}}{\alpha}x \right) \right]. \nonumber
\end{eqnarray}
The second term in Eq.~(\ref{Xi_xy}) is the difference between $\sigma(\Vec{x}_\tau,\Vec{y}_\tau)=\int_0^\tau \frac{dt}{T_{\lambda (t)}} \left[ -\dot{x} \circ (kx +f_{\lambda(t)} \alpha y) + -\dot{y} \circ (k_{\lambda(t)} \alpha^2 y  - f_{\lambda(t)}x/\alpha) \right ]$ and $\widetilde{\sigma}(\Vec{x}_\tau)=-\int_0^\tau \frac{dt}{T_{\lambda(t)}} [\dot{x} \circ kx]$, where $\circ$ represents the Stratonovich integral \cite{sekimoto_book2010}.
If $\lambda(t)$ is fixed in time,
we may show that $\Xi$ in the $\alpha \rightarrow \infty$ limit has the steady average rate,
\begin{eqnarray}
\left \langle\Xi(\Vec{x}_\tau, \Vec{y}_\tau) \right \rangle_{\lambda, \tau}/\tau \xrightarrow {\tau \to \infty}  f^2/\gamma k. \label{Xi_xy_SS}
\end{eqnarray}
Since this rate is positive, we find that the reduced dynamics written by Eq.~(\ref{xcg}) does not reproduce the entropy production in the dynamics (\ref{xymodel}). See \cite{nakayama2012} for the general analysis. 
Note that Eq.~(\ref{integral_anom}) holds not only for the trivial case where $\Xi = 0$, but also for $\langle \Xi\rangle > 0$ situations as in this example.

The sufficient condition for Eq.~(\ref{integral_anom}) to hold is
that the original dynamics (before coarse-graining) only contains
time-reversal invariant variables.
Here we note on the case of eliminating the momentum variable (which is not time-reversal invariant) from the underdamped Langevin dynamics.
We consider a one-dimensional model with position dependent temperature \cite{celani2012,nakayama2012},
\begin{eqnarray}
\begin{array}{l}
\dot{x} = p/m , \\
\dot{p} = -\gamma p/m + F_{\lambda(t)} (x) + \sqrt{2\gamma T_{\lambda(t)}(x)} \xi(t).
\end{array} \label{xpmodel}
\end{eqnarray}
with $p$ the momentum variable, $m$ the mass of the Brownian particle, $F(x)$ the general $x$ dependent force.
Since $\bar{p}=-p$, $\Xi(\Vec{x}_\tau, \Vec{p}_\tau)$ does not satisfy Eq.~(\ref{integral_anom}) for general parameters.
However, in the case where the parameters in Eq.~(\ref{xpmodel}) justify the elimination of $p$ and therefore the effective dynamics obeys the overdamped equation,
\begin{eqnarray}
\gamma \dot{x} = F_{\lambda(t)}(x) -\frac{\partial T_{\lambda(t)}(x)}{\partial x}+ \sqrt{2\gamma  T_{\lambda(t)}(x)} \circ \xi(t), \label{xpmodel_cg}
\end{eqnarray}
we may prove
\begin{eqnarray}
\left \langle e^{-\Xi \left(\Vec{x}_\tau,\Vec{p}_\tau \right)} \right \rangle_{\lambda, \tau} = 1. \label{integral_anom_xp} 
\end{eqnarray}
This specific equality was noted in \cite{celani2012}.
As in the previous example, we find that $\langle \Xi \rangle$ is positive for the general $\frac{\partial T_{\lambda(t)}(x)}{\partial x}\neq 0$ case \cite{hondou2000,celani2012,nakayama2012}.

\textit{Deterministic diffusion model.}---
Given the two examples we have shown, one might expect that the decrease of entropy production is a general consequence of the Markovian limit coarse-graining. 
In this section we clarify that this is not the case by considering
a deterministic Hamiltonian-like model.
In this model, a probabilistic dynamics could be derived in the appropriate coarse-graining limit.
Following the concept showed in the previous section, we formally define the entropy productions, and check that Eqs.~(\ref{integral_anom}), (\ref{ineq_anom}) are violated.
We find that the violation is due to the asymmetry $P_t(x,y)\neq P_t(\bar{x},\bar{y})$.
This is the second main claim of this paper.

\begin{figure}[htbp]
 \begin{center}
  \includegraphics[width=83mm]{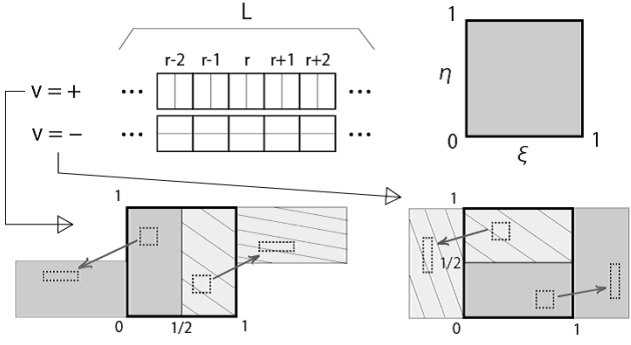}
 \end{center}
  \caption{\label{multibaker} Scheme of the multibaker map. The map considers the label of squares ($r$), two-dimensional coordinates inside the square ($\xi,\eta$), and the discretized velocity ($v$). 
The map is area-preserving and time-reversal invariant since the 
 transition rules in the $v=+$ and $v=-$ systems are exactly opposite to 
 each other [Eqs.~(\ref{transition_mb}), (\ref{trans_det2})].}
\end{figure}

Our Markov chain model is an extension of the multibaker map \cite{gaspard1992}.
The model is composed of many baker transformations that act on the nearest neighbor squares (Fig.~\ref{multibaker}).
The variables $\xi,\eta \in [0,1]$ are the coordinates inside each unit area squares, and $\boxr \in {1,2,...,L}$ is the label of those squares.
We set a periodic boundary condition for $\boxr$, and regard $\boxr=L+1$ as $\boxr=1$ and $\boxr=0$ as $\boxr=L$. 
We further introduce the ``discretized velocity'' variable, $v= +$ or $-$. 
The $v=+$ and $-$ systems are each composed of $L$ squares, and are considered to be separated and non-interacting. The 
transition probability (deterministic map) from $(\boxr,\xi,\eta,v)$ to $(\boxr',\xi',\eta',v')$ in a unit time step is written as,
\begin{eqnarray}
&&{\Trans}(\boxr',\xi',\eta',v'|\boxr,\xi,\eta,v) \nonumber\\
&& \ \ \ \ :=\begin{cases}
    {\Trans}^{+}(\boxr',\xi',\eta'|\boxr,\xi,\eta) & (v=v'=+) \\
    {\Trans}^{-}(\boxr',\xi',\eta'|\boxr,\xi,\eta) & (v=v'=-)\\
    0 & (\text{otherwise})
  \end{cases} \label{transition_mb}
\end{eqnarray}
with [see Fig.~\ref{multibaker}],
\begin{eqnarray}
&&{\Trans}^{+}(\boxr',\xi',\eta'|\boxr,\xi,\eta) := \theta \left(\tfrac{1}{2}-\xi \right) \delta_{\boxr',\boxr-1} \delta \left(\xi'-2\xi \right )\delta\left(\eta'-\tfrac{\eta}{2}\right)  \nonumber \\
&& \ \ \ \ \ \ +  \theta \left(\xi-\tfrac{1}{2} \right) \delta_{\boxr',\boxr+1} \delta \left(\xi'-2\xi+1 \right ) \delta  \left(\eta'-\tfrac{\eta}{2} - \tfrac{1}{2}\right) \label{trans_det1},\\
&& {\Trans}^{-}(\boxr',\xi',\eta'|\boxr,\xi,\eta) := \theta \left(\tfrac{1}{2}-\eta \right) \delta_{\boxr',\boxr+1} \delta \left(\xi'-\tfrac{\xi}{2} \right)\delta \left(\eta'-2\eta \right)  \nonumber \\
&& \ \ \ \ \ \ +  \theta \left(\eta-\tfrac{1}{2} \right) \delta_{\boxr',\boxr-1} \delta \left(\xi'-\tfrac{\xi}{2}-\tfrac{1}{2} \right ) \delta  \left(\eta'-2\eta +1 \right), \label{trans_det2}
\end{eqnarray}
where $\theta(\cdot)$ is the Heaviside step function and $\delta (\cdot)$ is the Dirac delta function.
Taking the time reversal of the variables as $(\bar{\boxr}, \bar{\xi}, \bar{\eta}, \bar{v})=(\boxr, \xi, \eta, -v)$, the 
model dynamics is completely time reversal symmetric, ${\Trans} (\boxr',\xi ',\eta ',v'|\boxr,\xi,\eta,v)={\Trans}(\boxr,\xi,\eta,-v|\boxr',\xi ',\eta ',-v')$.

Given an initial point $(\boxr_0,\xi_0,\eta_0,v)$, we gain a deterministic trajectory after $N$ time steps, $\left(\Vec{\boxr}_N,\Vec{\xi}_N,\Vec{\eta}_N,v \right)$.
Writing the probability density function of $(\boxr, \xi, \eta, v)$ at time $t$ as ${\Prob}^v_t(\boxr,\xi, \eta)$, the Shannon entropy difference after $N$ time steps is zero,
\begin{eqnarray}
s\left(\Vec{\boxr}_N,\Vec{\xi}_N,\Vec{\eta}_N,v \right) &=& \log \frac{{\Prob}^v_{0}(\boxr_0,\xi_0,\eta_0)}{{\Prob}^v_{N\Delta t}(\boxr_N,\xi_N,\eta_N)}=0,
 \label{shannon_mb}
\end{eqnarray}
because the phase space volume is conserved along the trajectories.
We also find
\begin{eqnarray}
\sigma\left(\Vec{\boxr}_N,\Vec{\xi}_N,\Vec{\eta}_N,v \right)&& = \log \frac{{\Trans}^v(\Vec{\boxr}_N,\Vec{\xi}_N,\Vec{\eta}_N|\boxr_0,\xi_0,\eta_0)}{{\Trans}^{-v}(\Vec{\boxr}^\dagger_N,\Vec{\xi}^\dagger_N,\Vec{\eta}^\dagger_N|\boxr_N,\xi_N,\eta_N)}\nonumber \\
&&=0, \label{heat_mb}
\end{eqnarray}
since we have set the dynamics in $v=-$ to be the complete time reversal of $v=+$.
Therefore we have confirmed $\Sigma \left (\Vec{\boxr}_N,\Vec{\xi}_N,\Vec{\eta}_N,v \right) = s\left(\Vec{\boxr}_N,\Vec{\xi}_N,\Vec{\eta}_N,v \right)+\sigma\left(\Vec{\boxr}_N,\Vec{\xi}_N,\Vec{\eta}_N, v \right) = 0$.
Note that this holds for any given initial density function ${\Prob}^v_0(\boxr_0,\xi_0, \eta_0)$.

Now we consider reducing the variables and only observing the discrete variable $\boxr$. Then, the reduced total entropy 
production $\widetilde{\Sigma}(\Vec{\boxr}_N)$  could be written using the ``mesoscopic'' probability distribution 
$\widetilde{\Prob}_t(\boxr) := \sum_{v} \int_0^1 d\xi \int _0^1 d\eta {\Prob}^v_t(\boxr,\xi,\eta)$, 
and transition probabilities
$\widetilde{\Trans}_t[\boxr'|\boxr,{\Prob}_0^{\pm}(\cdot)]:= \sum_{v} \int d\xi d\eta d\xi' d\eta' \Prob^v _t(\boxr,\xi,\eta) {\Trans}^v(\boxr',\xi',\eta'|\boxr,\xi,\eta)/\widetilde{\Prob}_t(\boxr)$,  $\widetilde{\Trans}^\dagger_t[\boxr'|\boxr,{\Prob}_{N\Delta t} ^{\pm}(\cdot)]:= \sum_{v} \int d\xi d\eta d\xi' d\eta' \Prob^{v \dagger} _{t}(\boxr,\xi,\eta) \times \newline{\Trans}^v(\boxr',\xi',\eta'|\boxr,\xi,\eta)/\widetilde{\Prob}^\dagger_{t}(\boxr)$.
We find from Eq.~(\ref{integral_cg}) that for general initial distributions, the following holds.
\begin{eqnarray}
 \left \langle \Xi \left(\Vec{\boxr}_N,\Vec{\xi}_N,\Vec{\eta}_N,v \right) \right \rangle_ N = - \left \langle \widetilde{\Sigma}(\Vec{\boxr}_N) \right \rangle_N  \leq  0. \label{ineq_anom2}
\end{eqnarray}
To confirm that the $\langle \widetilde{\Sigma}(\Vec{\boxr}_N) \rangle _N \neq 0$ case exists, we assume that ${\Prob}^{+} _0(\boxr,\xi,\eta)$ and ${\Prob}^{-} _0(\boxr,\xi,\eta)$ are smooth in the $\xi$ and $\eta$ direction, respectively, and define the time scale of the dynamics inside the squares as
\begin{eqnarray}
\tau_{\xi,\eta}:= \log \sup_{\boxr,\xi,\eta} \left\{ \left| \frac{\partial}{\partial \xi} {\Prob}^{+} _0(\boxr,\xi,\eta) \right|, 
\left| \frac{\partial}{\partial \eta} {\Prob}^{-} _0(\boxr,\xi,\eta) \right| \right\}.
\end{eqnarray}
Then we may take, if $L$ is sufficiently large, the ``mesoscopic time scale'' $t^*$, satisfying $\tau_{\xi,\eta} \ll t^* 
\ll \tau_\boxr$. Here, $\tau_\boxr $ $(\sim L^2)$ is the typical time for $\widetilde{P}_t (\boxr)$ to become uniform.
After this time $t^*$, the dynamics reduces to a simple random walk in the $\boxr$ direction,
\begin{eqnarray}
\widetilde{\Trans}_t, \widetilde{\Trans}^\dagger_t  \xrightarrow{t\geq t^*} \widetilde{\Trans}(\boxr'|\boxr) = \frac{1}{2}\delta_{\boxr',\boxr+1} + \frac{1}{2}\delta_{\boxr',\boxr-1}.
\end{eqnarray}
Retaking the initial time $t=0$ at this $t^*$, the average total entropy production of the reduced dynamics satisfies
\begin{eqnarray}
&&\left \langle \widetilde{\Sigma}(\Vec{\boxr}_N) \right \rangle _N = \sum_{\boxr_0,\boxr_1,...,\boxr_N} \widetilde{\Prob}_0(\boxr_0) \widetilde{\Trans}(\Vec{\boxr}_N|\boxr_0) \nonumber\\
&&\hspace{17mm}
\times\log \frac{\widetilde{\Prob}_0(\boxr_0)\widetilde{\Trans}(\Vec{\boxr}_N|\boxr_0)}{\widetilde{\Prob}_{N\Delta t}(\boxr_N)\widetilde{\Trans}(\Vec{\boxr}^\dagger_N|\boxr_N)}\geq 0. \label{I_ineq}
\end{eqnarray}
Equality in (\ref{I_ineq}) [and (\ref{ineq_anom2})] is achieved only when $\widetilde{\Prob}_0(\boxr_0)\widetilde{\Trans}(\Vec{\boxr}_N|\boxr_0)=\widetilde{\Prob}_{N\Delta t}(\boxr_N)\widetilde{\Trans}(\Vec{\boxr}^\dagger_N|\boxr_N)$ for all $\Vec{\boxr_N}$, that is, only when the given initial distribution is the equilibrium state, $\widetilde{\Prob}_0(\boxr)=1/L.$
Hence, we observe that the entropy production increases after the reduction in this model, as opposed to the case where Eq.~(\ref{ineq_anom}) holds.

The inequality (\ref{ineq_anom2}) states that the integral fluctuation theorem [Eq.~(\ref{integral_anom})] does not hold in this model, except for the trivial case $\widetilde{\Sigma}=0$ (equilibrium state).
From Eq.~(\ref{prof1}) in the following section, we notice that the violation of Eq.~(\ref{integral_anom}) is due to the broken symmetry in the density function, $P^{+}_{N\Delta t}(\cdot) \neq P^{-}_{N\Delta t}(\cdot)$, which is valid for any $N>0$ including $N\to \infty$ in this model.
This is in clear contrast with the underdamped Langevin model, where the symmetry emerges in the overdamped (Markovian) limit [see Eq.~(\ref{prof3})].

\textit{Proofs.}---
First we see that
\begin{eqnarray}
&&\left \langle e^{-\Xi\left(\Vec{x}_N,\Vec{y}_N \right)} \right \rangle_{\lambda, N} = \int d\Vec{x}_N d\Vec{y}_N {\Prob}_{\lambda}(\Vec{x}_N,\Vec{y}_N) \nonumber \\
&& \times
\frac{{\Prob}_{N\Delta t}(x_N,y_N) {\Trans}_{\lambda^\dagger} (\Vec{x}^\dagger_N,\Vec{y}^\dagger_N|\bar{x}_N,\bar{y}_N)\widetilde{\Prob}_{0}(x_0) \widetilde{\Trans}_{\lambda}[\Vec{x}_N|x_0]}
{{\Prob}_{0}(x_0,y_0) {\Trans}_{\lambda} (\Vec{x}_N,\Vec{y}_N|x_0,y_0)\widetilde{\Prob}_{N\Delta t}(x_N) \widetilde{\Trans}_{\lambda^\dagger}[\Vec{x}^\dagger_N|\bar{x}_N]} \nonumber \\
&& \ \  =\int d\Vec{x}_N d\Vec{y}_N {\Prob}_{\lambda^\dagger}(\Vec{x}^\dagger_N,\Vec{y}^\dagger_N)
\frac{{\Prob}_{N\Delta t}(x_N,y_N)}{{\Prob}_{N\Delta t}(\bar{x}_N,\bar{y}_N)} \nonumber \\
&& \hspace{20mm}\times
\frac{\widetilde{\Prob}_{0}(x_0) \widetilde{\Trans}_{\lambda}[\Vec{x}_N|x_0]}
{\widetilde{\Prob}_{N\Delta t}(x_N) \widetilde{\Trans}_{\lambda^\dagger}[\Vec{x}^\dagger_N|\bar{x}_N]}.
 \label{prof1}
\end{eqnarray}
For simplicity we omitted $P_0, P_{N \Delta t}$ in $\widetilde{\Trans}_{\lambda}, \widetilde{\Trans}_{\lambda^\dagger}$, respectively.
Now if $\bar{x}=x$, $\bar{y}=y$, the right hand side of Eq.~(\ref{prof1}) reduces to,
\begin{eqnarray}
&&\int d\Vec{x}_N d\Vec{y}_N {\Prob}_{\lambda^\dagger}(\Vec{x}^\dagger_N,\Vec{y}^\dagger_N)
\frac{\widetilde{\Prob}_{0}(x_0) \widetilde{\Trans}_{\lambda}[\Vec{x}_N|x_0]}{\widetilde{\Prob}_{N\Delta t}(x_N) \widetilde{\Trans}_{\lambda^\dagger}[\Vec{x}^\dagger_N|x_N]}\nonumber \\
&& \ \ \ = \int d\Vec{x}_N \widetilde{\Prob}_{0}(x_0) \widetilde{\Trans}_{\lambda}[\Vec{x}_N|x_0] = 1,  \label{prof2}
\end{eqnarray}
which is Eq.~(\ref{integral_anom}). In the case of the underdamped model (\ref{xpmodel}), for a small parameter $\epsilon:= \tau_p/ \tau_x$ ($\tau_x$ is the fastest time scale of the motion in the $x$ direction, and $\tau_p := m/\gamma$), we may show that the ratio $P_{\tau} (x, p)/P_{\tau} (x, -p)$ becomes close to unity, $1+O(\epsilon)$, assuming that $\tau$ is large enough compared to $\tau_p$. Since $\epsilon \to 0$ corresponds to the overdamped limit, Eq.~(\ref{prof1}) is now,
\begin{eqnarray}
&&\left \langle e^{-\Xi \left( \Vec{x}_\tau,\Vec{p}_\tau \right)} \right \rangle_{\lambda, \tau} = \int d\Vec{x}_\tau d\Vec{p}_\tau {\Prob}_{\lambda^\dagger}(\Vec{x}^\dagger_\tau,\Vec{p}^\dagger_\tau) \nonumber \\
&&\hspace{5mm}\times \frac{\widetilde{\Prob}_{0}(x_0) \widetilde{\Trans}_{\lambda}[\Vec{x}_\tau|x_0]}{\widetilde{\Prob}_{\tau}(x_\tau) \widetilde{\Trans}_{\lambda^\dagger}[\Vec{x}^\dagger_\tau|x_\tau]}[1+O(\epsilon)] 
\xrightarrow{\epsilon \to 0}  1,  \label{prof3}
\end{eqnarray}
which is Eq.~(\ref{integral_anom}).

\textit{Remarks and conclusion.}---
Recently, entropy production from accessible degrees of freedom was experimentally measured and analysed in \cite{mehl2012}.
Although their definition of ``apparent entropy production'' is different to our coarse-grained entropy production, the heat [Eq.~(\ref{micro_heat_cg})] and the hidden entropy productions [Eq.~(\ref{def_anom})] are also measurable quatities in their experimental setup.
Therefore, we claim that our main result Eq.~(\ref{integral_anom}) can be experimentally tested.

Next, we note the relation between our study and recent results on steady state thermodynamics.
We observe \cite{nakayama2012} that $\widetilde{\sigma}(\Vec{x})$ in the $\alpha \to \infty$ limit of the model (\ref{xymodel}) is equivalent to Hatano-Sasa's excessive entropy production \cite{hatanosasa2001}.
In this scheme, $\Xi(\Vec{x},\Vec{y})$ is the so-called housekeeping entropy production, therefore Eq.~(\ref{integral_anom}) is equivalent to the integral fluctuation theorem shown in \cite{speck2005}. It is similar in the case of the model (\ref{xpmodel}) if (\ref{xpmodel_cg}) presents equilibrium dynamics \cite{nakayama2012}, thus in this case Eq.~(\ref{integral_anom_xp}) is the fluctuation theorem for the housekeeping heat extended to the underdamped Langevin model \cite{spinney2012}.

In conclusion, we have shown through various examples that the entropy production may in general increase or decrease after reducing variables.
For a generic class of systems with only time-reversal invariant variables,
and for special cases where the final density is symmetric for all time-reversal anti-symmetric variables,
Eq.~(\ref{integral_anom}) is satisfied and the entropy production always decreases.
In the deterministic diffusion model, where the entropy production increases, Eq.~(\ref{integral_anom}) is violated due to the asymmetric density function.
Since Eq.~(\ref{integral_anom}) has to be violated for the total entropy production to increase after reducing variables, 
we find that such asymmetry of the distribution function is crucial in the general derivation of stochastic processes from Hamiltonian dynamics.
It is left for future studies to clarify this point in more physical kinetic equations.

\textit{Acknowledgments.}---
We thank T.~Sagawa, M.~Sano, S.-i.~Sasa, and  K.~A.~Takeuchi for fruitful discussions and reading of the manuscript. This work was supported by JSPS research fellowship.

%\bibliography{2012kawaguchi}% Produces the bibliography via BibTeX.

%

\end{document}